\documentclass{aa}
\usepackage{times}
\usepackage{graphics}
\begin{document}

\thesaurus{22(03.09.6; 03.19.2; 08.09.2 HD 93521; 13.21.2)}

\title{The ORFEUS\,II  Echelle spectrometer:\\
Instrument description, performance and data reduction}
\author{
J.\,Barnstedt\inst{1} \and
N.\,Kappelmann\inst{1} \and
I.\,Appenzeller\inst{2} \and
A.\,Fromm\inst{1} \and
M.\,G\"{o}lz\inst{1} \and
M.\,Grewing\inst{1,3} \and
W.\,Gringel\inst{1} \and
C.\,Haas\inst{1} \and
W.\,Hopfensitz\inst{1} \and
G.\,Kr\"{a}mer\inst{1} \and
J.\,Krautter\inst{2} \and
A.\,Lindenberger\inst{1} \and
H.\,Mandel\inst{2} \and
H.\,Widmann\inst{1}}

\offprints{barnstedt@astro.uni-tuebingen.de}

\institute{Institut f\"{u}r Astronomie und Astrophysik, Abt. Astronomie, Eberhard-Karls-%
Universit\"{a}t T\"{u}bingen, Waldh\"{a}userstr. 64, D-72076 T\"{u}bingen, Germany \and
Landessternwarte Heidelberg, K\"{o}nigstuhl, D-69117 Heidelberg, Germany \and
Institut de Radio Astronomie Millim\'{e}trique (IRAM), 300 Rue de la Piscine, F-38406 Saint
Martin d'H\`{e}res, France}

\titlerunning{The ORFEUS\,II  Echelle Spectrometer}

\date{Received May 20; accepted September 15, 1998}

\maketitle

\begin{abstract}

During the second flight of the ORFEUS-SPAS mission in
November/December 1996, the Echelle spectrometer was
used extensively by the Principal and Guest
Investigator teams as one of the two focal plane
instruments of the ORFEUS telescope.
We present the in-flight performance and the principles of
the data reduction for this instrument. The wavelength
range is 90\,nm to 140\,nm, the spectral resolution
is significantly better than
$\lambda$/$\Delta$$\lambda$\,=\,10\,000, where
$\Delta$$\lambda$ is measured as FWHM of the
instrumental profile. The effective area peaks at
1.3\,cm$^{2}$ near 110\,nm. The background is
dominated by straylight from the Echelle grating and
is about 15\% in an extracted spectrum for spectra
with a rather flat continuum. The internal accuracy of
the wavelength calibration is better than
$\pm$\,0.005\,nm.

\end{abstract}

\section{Introduction}

The German spacecraft ORFEUS-SPAS was the primary
payload aboard the Space Shuttle {\it Columbia\/}
during mission \mbox{STS-80}, flown from 19 November to 7
December, 1996. The ORFEUS telescope was the main
instrument aboard the {\bf astro}nomy platform {\bf
S}huttle {\bf Pa}llet {\bf S}atellite (ASTRO-SPAS).

The ASTRO-SPAS is a free flying platform that is
designed to operate autonomously in the vicinity (up
to 60\,km) of the Space Shuttle for a limited period
of up to 14 days. It was designed by the
Daimler-Benz Aerospace AG (DASA) in Ottobrunn,
Germany.

With a mission duration of 17.7 days STS-80 was the
longest shuttle flight to date, the free flying time
of ORFEUS-SPAS was 14 days. ORFEUS-SPAS was released a
few hours after launch on 20 November, 1996. It was
the third flight of the reusable ASTRO-SPAS platform
and the second flight of the ORFEUS-SPAS
configuration.

The ORFEUS telescope ({\bf O}rbiting and {\bf
R}etrievable {\bf F}ar and {\bf E}xtreme {\bf
U}ltraviolet {\bf S}pectrometer) (Kr\"{a}mer et al. \cite{kraemer1},
\cite{kraemer2}, Grewing et al. \cite{grewing1},
\cite{grewing2}) consists of a
1\,m normal incidence mirror with a focal length of
2.4\,m and two spectrometers as focal plane
instrumentation: the Berkeley spectrograph and the
Echelle spectrograph. The Berkeley spectrograph was
developed by the Space Sciences Laboratory of the
University of California at Berkeley
(Hurwitz \& Bowyer \cite{hurwitz1},
\cite{hurwitz2}, Hurwitz et al. \cite{hurwitz3}) and the
Echelle spectrograph was developed by the University
of T\"{u}bingen and the Landessternwarte Heidelberg
(Appenzeller et al. \cite{appenzeller}). As a third independent
instrument, the Interstellar Matter Absorption Profile
Spectrometer (IMAPS) was attached to the ASTRO-SPAS
(Jenkins et al. \cite{jenkins}).

The ORFEUS telescope and the Echelle spectrometer were
designed and built by Kayser-Threde GmbH in Munich,
Germany. The Echelle detector as well as the Echelle
electronics and the Echelle onboard processor were
designed and built by the Institute for Astronomy and
Astrophysics, Department Astronomy, of the University
of T\"{u}bingen.

The ORFEUS-SPAS total instrument operation time was
263\fh0, while the net integration time was 164\fh9,
resulting in an efficiency of 62.5\% for all
instruments. This is an extremely high value as
compared to other satellite missions.

50\% of the total integration time were reserved for
Guest Investigators selected by peer review. The other
50\% were used by the Principal Investigator teams who
had provided the instruments. 65 distinct
objects were observed with the Echelle spectrometer.

In this paper we present the performance figures for
the Echelle spectrometer and a short description of the
principles of the data extraction procedure.

\section{Instrument description}

\begin{figure}[tbp]

\resizebox{\hsize}{!}{\includegraphics{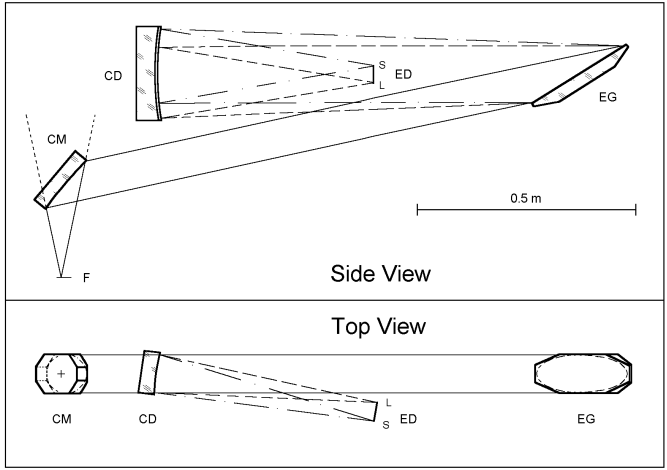}}

\caption{Optical design of the Echelle spectrometer.
Legend: F\,= telescope focus and entrance diaphragm,
CM\,= collimator mirror, EG\,= Echelle grating, CD\,=
cross disperser, ED\,= Echelle detector, L\,= long
wavelength side, S\,= short wavelength side.}

\label{fig1}
\end{figure}

Within the ORFEUS telescope a movable mirror is used
to switch between the two spectrometers. We will
restrict the instrument description to the Echelle
spectrometer.

The movable mirror is designed to act at the same time
as an off-axis parabolic collimator (CM, Fig.~\ref{fig1}).
When moved in it feeds the Echelle grating (EG) with a
parallel light bundle. When the collimator mirror is
moved out, the beam is falling directly into the
Berkeley spectrometer.

The Echelle grating has a groove density of 316
lines/mm with a blaze angle of 62\fdg5. It is operated
in the diffraction orders 40 to 61 covering the
wavelength range from 90\,nm to 140\,nm.

The Echelle diffraction orders are separated by means
of a spherical cross disperser grating (CD) with 1200
lines/mm which also serves as the spectrograph camera.
This design minimizes the number of reflections,
which, in view of the low reflectivities in the FUV
(ranging from 10\% to 30\%), was of critical
importance for achieving an acceptable overall
efficiency.

The spectrum is recorded by the Echelle detector (ED),
which has a plane surface and a sensitive area of
40\,mm\,$\times$\,40\,mm.

The Echelle spectrometer is designed to achieve a
spectral resolution of
$\lambda$/$\Delta$$\lambda$\,=\,10\,000 when used with
an entrance aperture of 10{\arcsec} diameter
(Appenzeller et al. \cite{appenzeller}).

\subsection{The Echelle Detector}

The Echelle detector is a photon counting microchannel
plate detector with a wedge and strip readout system.
Three stacked microchannel plates (MCPs), operated in
a Z-configuration, provide a gain of 10$^{7}$ to
10$^{8}$ electrons/photon. Each of the three MCPs has
its own power supply and an accelerating voltage is
applied to the gaps between the MCPs. During flight
the gain of the Echelle detector was monitored and
could be controlled by variation of the high voltage
of the third MCP. The high voltage could be adjusted
by commands during flight or by starting an automatic
function that adjusted the high voltage so that a
predefined gain value was achieved.

A repeller grid in front of the MCPs produces an electric
field of about 50\,V/mm, which is used to force those photo
electrons back into the MCP channels, which are
released from the areas in between the channels. This
improves the quantum efficiency by about 30\% but also causes
a loss of 10\% due to the shading by the repeller
grid.

A four electrode wedge-and-strip anode (Martin et al.
\cite{martin})
was used as readout system for the MCP detector.
A special design allowed for a
significant reduction of the edge distortions, which
are basically unavoidable. Further, by oversizing the
active area of the anode to 44\,mm\,$\times$\,44\,mm,
it was possible to avoid any distortions within the
sensitive area of the detector.
The image format of the detector is 1024 pixels by 512
pixels, which corresponds to the active area of the anode.

The detector sends for
each photon event x/y-coordinates to the Echelle onboard
processor where these events are integrated onto an
image memory. The content of this image memory is
stored on magnetic tape at the end of each
observation.

In parallel, the x/y-coordinates for each single
photon event are stored on tape directly, thus keeping
also the photon arrival time to an accuracy of better
than 1 second. This method is limited by the ASTRO-%
SPAS interface to an event rate of less than 3500
counts/s, whereas the onboard integration works with
much higher photon rates up to 30\,000 counts/s. The
electronic dead time is about 13\,$\mu$s per event.

The count rate of the Echelle detector is monitored
online and also stored on tape together with many
other housekeeping data. This count rate is derived
from all events registered by the charge amplifiers
falling above the lower threshold of the detector
electronics. Not all of these events are finally
integrated into the image for the following reasons:
\begin{itemize}
\item An upper electronic threshold suppresses pulses
too large to be processed by the position decoding
electronics.
\item At high count rates the dead time of 13\,$\mu$s
for position decoding reduces the rate of registered
events accordingly.
\item Dead time effects from the interface electronics
and from the onboard processor further reduce the
electronic efficiency.
\end{itemize}

These effects are the cause that in general the rate of the
integrated counts within an image is between 15\% and
25\% lower than the count rate of the lower electronic
threshold.

In the main dispersion direction somewhat more than 3
pixels correspond to one optical resolution element of
$\Delta$$\lambda$\,=\,$\lambda$/10\,000 (161\,$\mu$m).
The electronic resolution was estimated to be about
1.5 detector pixels FWHM. So the electronic detector
resolution is sufficient to maintain the optical
resolution of the Echelle spectrometer.

Electronic test pulses were fed onto the
anode with a frequency of 10\,Hz at two different detector
positions and with 3 different pulse heights. These test pulses
were extremely useful during the checkout of the instrument
and provide a means of monitoring the electronic performance
of the detector throughout the mission.

\section{Extraction of Echelle orders}

The extraction of the Echelle spectra from the Echelle
detector images in x-direction (main
dispersion direction) is done by summing up 11 pixels (orders
40 to 49) or 9 pixels (orders 50 to 61) in y-%
direction. The center of the extraction in y-direction
follows a straight line along each order, but is
located on integer pixel numbers.

Some Echelle images show tilted absorption lines
within the strip of the Echelle orders. The reason for
this effect is still unclear. In such cases the
extraction was done summing up the pixels tilted by 45
degrees, which results in a significant increase in
resolution. However an additional wavelength error was
introduced by this procedure since the observed
wavelength in this way became dependent on the
centering of the Echelle order within the extraction
strip. This is a nonlinearity error of the size of
one pixel, which corresponds to 1/3 of the optical
resolution element and is equal to
$\Delta$$\lambda$\,=\,$\lambda$/30\,000.

\subsection{Background}

Between the Echelle orders a 3 pixel wide area was
used to estimate the background. With exception of
orders 40, 41 and 42 the background was calculated as
average of the strip above and below the corresponding
order. For the first three orders only the background
values below each order were calculated. The so
calculated background was smoothed prior to
being used for subtraction.

The intrinsic background of the detector was about
10$^{-5}$\,counts/s/pix. As 9 or 11 detector pixels
are added to one spectral pixel, this corresponds to
about 10$^{-4}$\,counts/s/pix in the extracted
spectra, which is negligible.

By far the strongest contribution to the background is
straylight from the Echelle grating. The contribution
in counts/detector-pixel is about 5\% to 10\% for the
low Echelle orders and 10\% to 20\% for the high
Echelle orders as these are closer together. The
percentage values refer to the maximum
counts/detector-pixel values of the Echelle orders.
The contribution in counts/spectrum-pixel is about
15\% for spectra with a rather flat continuum.

The Echelle straylight is scattered exactly
horizontally across the detector, whereas the spectral
orders run slightly tilted across the detector
corresponding to the dispersion of the cross
disperser. Hence emission lines produce additional
straylight while broad absorption lines lead to a
reduced straylight contribution in a horizontal line
at the y-position of this absorption line. These
varying straylight contributions are not yet correctly
handled by the background subtraction algorithm.

An additional contribution is caused by particle
events related to the South Atlantic Anomaly (SAA).
These produced count rates of up to 1000 counts/s
resulting in up to 2$\cdot$10$^{-3}$ counts/s/pix on
the detector and 2$\cdot$10$^{-2}$ counts/s/pix in the
spectra. The maximum duration of the passage through
the SAA was about 12\,min and the count rate curve
showed a quasi Gaussian shape during this time. In the
case of weak targets where the SAA background reduces
the signal to noise ratio of the spectrum
significantly, we are able to integrate the spectrum
from the single photon records excluding the SAA
transit period.

\begin{figure}[tbp]

\resizebox{\hsize}{!}{\includegraphics{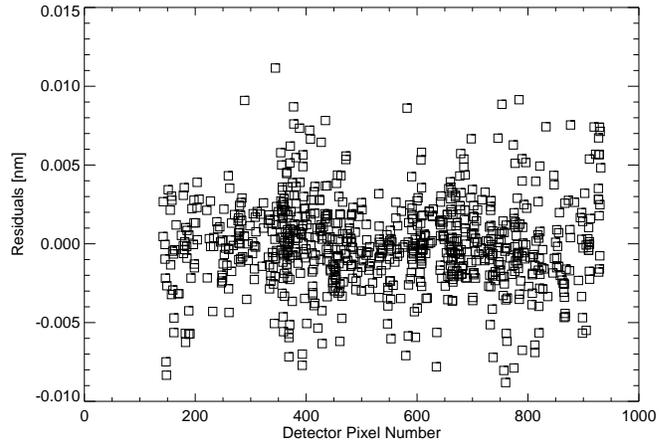}}

\caption{Residuals of the wavelength calibration fit.
814 positions of interstellar absorption lines from 12
different objects were used to fit 7 wavelength
calibration parameters.}

\label{fig2}
\end{figure}

\subsection{Correction of the blaze function}

The efficiency of blazed Echelle gratings varies as
(sin(x)/x)$^{2}$ (blaze function)
(Schroeder \& Hiliard \cite{schroeder}),
where x depends on the deviation of the
output angle from the specular angle of maximum
efficiency (blaze angle) and on the wavelength. The
optimized diffraction direction was preflight adjusted
to the y-centerline of the Echelle detector. In the
flight data we found the center of maximum efficiency
to differ between the Echelle orders and furthermore
both the position and the width of the blaze function
differing from observation to observation. Therefore
it was necessary to introduce an individual blaze
correction for each observation. For spectra that
showed a relatively undisturbed continuum the blaze
correction could be done with a rather high accuracy.
However for many spectra which are dominated by
absorption lines only the overlap region between two
adjacent orders could be used as a criterion for a
good blaze correction, resulting in a less reliable
correction.

After completion of the blaze correction for all
observations no systematic time dependent effects in
the variation of the blaze function were found.

\subsection{Wavelength calibration}

The wavelength calibration was calculated from the
positions of 814 interstellar absorption lines in the
Echelle images of 12 different objects. The centers of
the lines were determined by fitting Gaussian functions
to the profiles using a preliminary wavelength
calibration. After identification of the lines a table
with laboratory wavelengths and corresponding detector
pixel numbers was created. Using the preliminary
wavelength calibration, a mean radial velocity
component for each object was estimated and corrected
for. This approach assumes, that for each individual
object all identified interstellar absorption lines,
mainly H$_{2}$ and neutral elements, originate in the
same volume of gas.

\begin{figure}[tbp]

\resizebox{\hsize}{!}{\includegraphics{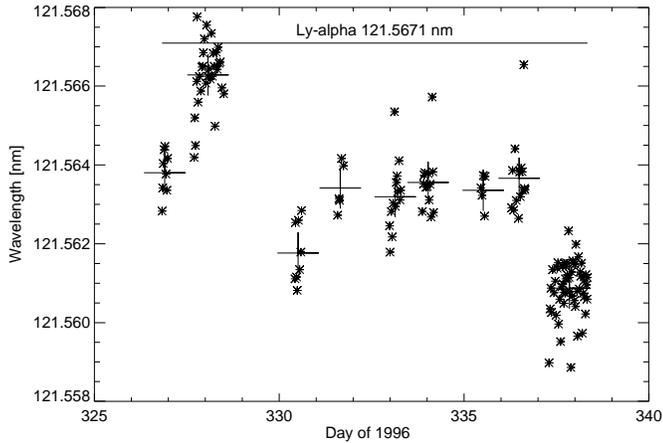}}

\caption{Apparent wavelengths of the geocoronal Ly-%
$\alpha$ emission line plotted versus time of
observation. For each of the 9 observing blocks for
the Echelle spectrometer the large crosses mark the
average Ly-$\alpha$ position, which was later used for
correction of the time dependent wavelength shift. For
these measurements no wavelength corrections except
for the satellite's orbital velocity were applied.}

\label{fig3}
\end{figure}

With a least square fit algorithm we then determined a
new set of pixel to wavelength
conversion parameters.

In Fig.~\ref{fig2} the residuals for all lines demonstrate
that the mean accuracy of the wavelength
correspondence is better than $\pm$\,0.005\,nm, i.e.
better than the optical resolution of the instrument.

The above procedure results in a relative wavelength
calibration. In order to determine an absolute scale
we used the position of the geocoronal Ly-$\alpha$
emission line as a reference. We found that for
different observing blocks the position of the Ly-%
$\alpha$ line changed by up to 0.006\,nm (Fig.~\ref{fig3}).
The reason for this shift is still unclear, we assume
that it emerges from temperature drifts of the
telescope and the Echelle spectrometer. A correction
for this effect has been applied in each observing
block.

Shifting of the Echelle spectrum on the detector
surface (e.g. displacement of the target within the
aperture) basically results in a relative wavelength
shift and thus in a radial velocity shift rather than
in a wavelength shift. The reason is, that the
wavelength dispersion is proportional to the
wavelength.

The wavelength scale of the targets were shifted to
correct for radial velocity components due to the
Earth's movement (heliocentric correction) and due to
the satellite's orbital movement.

A further shift occurs when the target image is not
exactly centered within the 20{\arcsec} diaphragm. The
maximum resulting uncertainty due to the 10{\arcsec}
radius of the diaphragm is $\pm$\,1.2$\cdot$10$^{-4}$
as a relative wavelength error (corresponding to
$\pm$\,36\,km/s). This error was not corrected up to
now.

\begin{figure}[tbp]

\resizebox{\hsize}{!}{\includegraphics{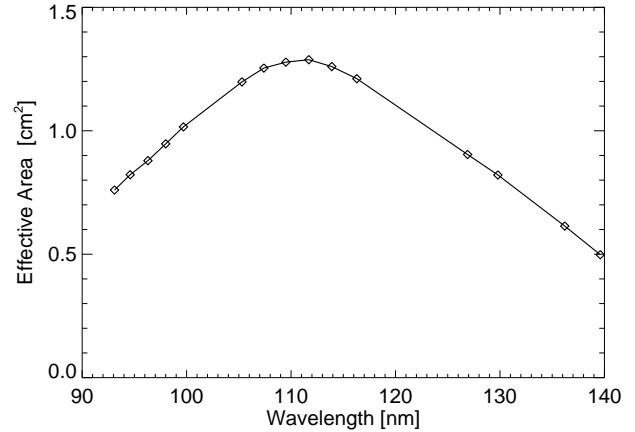}}

\caption{Effective area curve for the Echelle
spectrometer. The symbols denote the center
wavelengths of those Echelle orders for which a fit
value to the HST model spectrum was estimated.}

\label{fig4}
\end{figure}

\begin{figure*}[tbp]

\resizebox{\hsize}{!}{\includegraphics{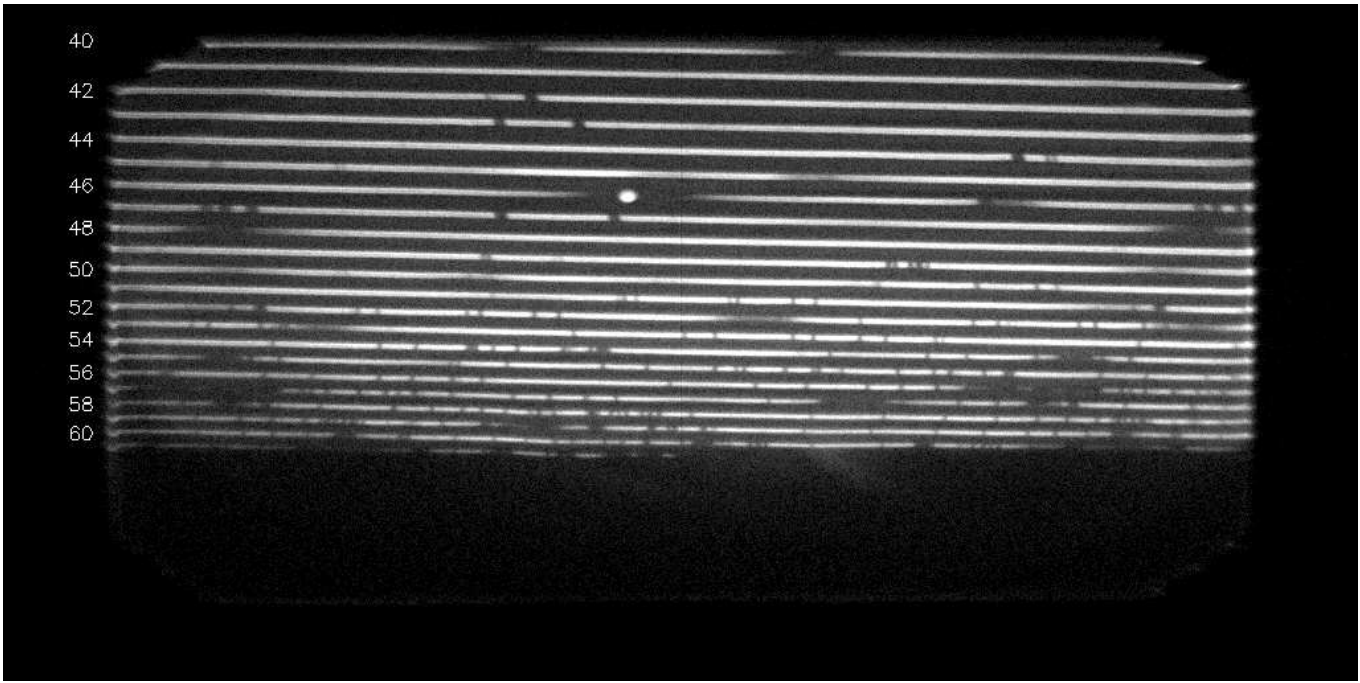}}

\caption{Echelle image of HD\,93521. The numbers at
the left side denote the Echelle orders. The image is
shown in the electronic format with
1024\,$\times$\,512 pixels, which does not correspond
to the square format of the detector of
40\,mm\,$\times$\,40\,mm. The Echelle spectrometer
could show some more Echelle orders than the visible orders 40
to 61, but below the Lyman limit at 91.15\,nm all
light is generally absorbed by interstellar hydrogen, so that
none of the spectra contains any useful flux below
this limit. (Orientation: long wavelengths are at top
and at left side).}

\label{fig5}
\end{figure*}

\subsection{Correction of loss of sensitivity in the
detector edges}

In the corners of the detector area and at the left edge
the efficiency of the repeller grid is reduced,
probably due to an inhomogenous field close to
the edge of the detector. A step occurs between
lower and normal sensitivity. In some images a
circularly shaped area is visible due to this effect. We
estimated a loss of about 25\% and corrected this by
applying a ``smooth'' step function. The position,
width and height of the step was estimated for each
order from the sum of all Echelle measurements.

A detailed flat field correction was not applied, due
to the fact, that the optical light path of the
spectrometer cannot be reproduced in our laboratory.
This however would be essential for an exact
estimation of the flat field behavior of the detector.
Any other correction methods are too uncertain to be
useful.

\subsection{Absolute flux calibration}

We used an HST archive model of G191\,B2B ``http://\-%
www.stsci.edu/\-ftp/\-cdbs/\-calspec/\-%
g191b2b\_mod\_002.tab'' as a reference for the
absolute flux calibration. The calibration was
crosschecked with a model of \object{BD\,+28\degr\,4211}
(R.\,Napiwotzki, priv. comm.). We estimate an accuracy of
$\pm$\,10\% for the flux calibration, provided the
object was fully centered within the aperture, which was not
always the case.

\section{Performance}

\subsection{Sensitivity}

The total efficiency of the Echelle spectrometer was
estimated before the flight by measuring the
efficiencies of the individual optical components.
The product of these values resulted in a maximum
effective area of 2.4\,cm$^{2}$.

For postflight recalibration we used a HST calibration
spectrum of \object{G191\,B2B}. By comparing the fluxes within
complete orders we estimated an effective area value
for the central wavelength of each Echelle order.
Values for Echelle orders that were contaminated by
strong absorption lines (e.g. Ly-$\alpha$) were
deleted from the data set. This procedure leads to a
maximum effective area of 1.3\,cm$^{2}$ near 110\,nm
(Fig.~\ref{fig4}), which is based on all detector
counts above the lower electronic threshold.

The 20{\arcsec}
diameter of the diaphragm should be sufficient to
maintain the total flux of the object, as the jitter
of the ASTRO-SPAS was about $\pm$\,2{\arcsec}. But we
have indications that the telescope was not precisely
focused during some periods of the mission, due to thermal
problems
during unexpected flight modes of the ASTRO-SPAS.
Together with the fact that some objects were not
fully centered this led to
significant nonstatistical variations of the count
rates, which resulted in a reduction of the flux and
thus sensitivity. Therefore the flux calibration was calculated for
the maximum observed count rate for the corresponding
target (also from other observations of this target,
if necessary). This count rate was scaled with the
registered count rate in the integrated image.

\subsection{Spectral resolution}

The spectrum of \object{HD\,93521} with a signal to noise ratio of
20 to 30 is best suited to study the
spectral resolution of the Echelle spectrograph.
It shows many interstellar absorption lines,
mainly from molecular hydrogen, which are very sharp
and rather strong. Among a total of 814 interstellar
line positions used for wavelength calibration, 159
line positions originated from this object. We coadded
two observations of this object with integration
times of 1080\,s and 660\,s respectively, taken within
two successive orbits (Fig.~\ref{fig5}).

\begin{figure}[tbp]

\resizebox{\hsize}{!}{\includegraphics{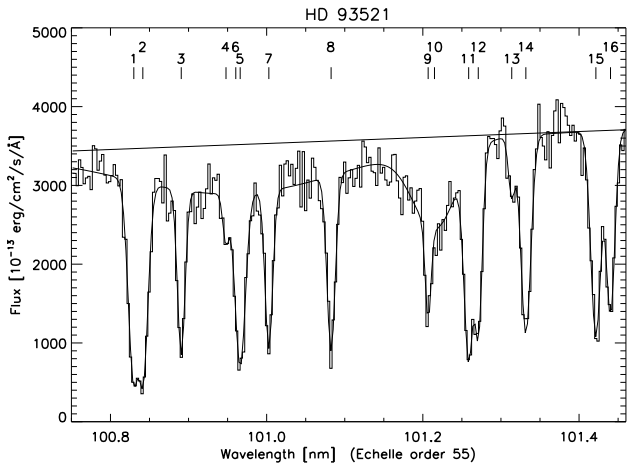}}

\caption{Part of the Echelle spectrum of HD\,93521.
This demonstrates the resolution achieved with the
Echelle spectrometer. The spectrum is plotted without
any smoothing. The smooth curve plotted over the
spectrum is a fit with several Gaussian shaped
absorption profiles. The straight line shows the
continuum used by the fitting algorithm. The continuum
is corrected by additional broad absorption profiles
(6 and 10). See Table 1 for the parameters of the
fitted curves.}

\label{fig6}
\end{figure}

We fitted many of these absorption lines in the
wavelength range from 96\,nm to 110\,nm with Gaussian
profiles and calculated the ratio of wavelength to the
FWHM of these lines. We got values mainly between
$\lambda$/$\Delta$$\lambda$\,=\,7000 to 10\,000 with
depths of 60\% to 80\% (Fig.~\ref{fig6} and Table\,1).

\begin{table*}[tbp]

\caption[]{Fitted Gaussian absorption profiles from
the spectrum shown in Fig.~\ref{fig6}. From the estimated
equivalent widths a Gaussian instrument profile was
calculated under the assumption that the original
absorption line had an depth of 100\%. The broad
stellar features are fitted for a better reproduction
of the continuum, as the fitting algorithm used
supports only a linear continuum.}

\begin{tabular}{rlllllrllr}
\hline
No. & $\lambda$ & Identification & Shift & Depth &
$\Delta$$\lambda$ & $\lambda$ / $\Delta$$\lambda$ &
Equival. & $\Delta$$\lambda$ instr. & $\lambda$ /
$\Delta$$\lambda$\\
 &  &  &  &  & (FWHM) &  & Width & (FWHM) & (instr.)\\
 & [nm] &  & [nm] & [\%] & [nm] &  & [nm] & [nm] & \\
\hline
1 & 100.830 & H$_{2}$ 100.8392 & -0.009 & 82.4 &
0.01653 & 6099 & 0.01449 & 0.00937 & 10755\\
2 & 100.841 & H$_{2}$ 100.8553 & -0.014 & 82.1 &
0.01538 & 6556 & 0.01343 & 0.00879 & 11470\\
3 & 100.891 & H$_{2}$ 100.903 & -0.012 & 71.1 &
0.01150 & 8775 & 0.00869 & 0.00809 & 12470\\
4 & 100.948 & Fe II 100.9580 & -0.010 & 21.4 & 0.00926
& 10907 & 0.00210 & 0.00904 & 11164\\
5 & 100.966 & H$_{2}$ 100.9772 & -0.011 & 76.2 &
0.01561 & 6469 & 0.01265 & 0.01012 & 9981\\
6 & 100.960 & stellar feature &  &  &  &  &  &  & \\
7 & 101.003 & H$_{2}$ 101.0132 & -0.010 & 68.3 &
0.01155 & 8756 & 0.00840 & 0.00843 & 11978\\
8 & 101.083 & H$_{2}$ 101.0941 & -0.011 & 71.2 &
0.01243 & 8133 & 0.00942 & 0.00873 & 11583\\
9 & 101.207 & H$_{2}$ 101.2173 & -0.010 & 48.0 &
0.00854 & 11845 & 0.00436 & 0.00750 & 13499\\
10 & 101.214 & stellar feature &  &  &  &  &  &  & \\
11 & 101.259 & H$_{2}$ 101.2681 & -0.009 & 76.3 &
0.01395 & 7256 & 0.01132 & 0.00903 & 11217\\
12 & 101.271 & H$_{2}$ 101.2822 & -0.011 & 67.3 &
0.01176 & 8611 & 0.00843 & 0.00869 & 11648\\
13 & 101.314 & Fe II 101.3262 & -0.012 & 20.9 &
0.00966 & 10490 & 0.00214 & 0.00945 & 10726\\
14 & 101.332 & H$_{2}$ 101.3434 & -0.011 & 69.5 &
0.01507 & 6726 & 0.01114 & 0.01084 & 9350\\
15 & 101.422 & H$_{2}$ 101.4334 & -0.011 & 70.7 &
0.01521 & 6669 & 0.01144 & 0.01075 & 9431\\
16 & 101.440 & H$_{2}$ 101.4509 & -0.011 & 63.4 &
0.01306 & 7765 & 0.00881 & 0.01010 & 10040\\
\hline
\end{tabular}
\end{table*}

We further estimated an upper limit for the width of
the instrument profile: Assuming that the absorption
lines had originally depths of 100\% with the same
equivalent width, we calculated the width of the
instrument profile that would be necessary to
reproduce the observed absorption line profiles. For
simplicity we assumed Gaussian profiles for both the
absorption lines and the instrumental profile and a
quadratic addition of the widths. The resulting values
for the FWHM of the instrumental profiles are in the
range $\lambda$/$\Delta$$\lambda$\,=\,9000 to 13\,000
(Table\,1). As these calculations are estimates of a
lower limit of the resolution of the Echelle
spectrometer, we conclude that the achieved resolution
of the Echelle spectrometer was significantly better
than the goal of
$\lambda$/$\Delta$$\lambda$\,=\,10\,000.

\begin{figure}[tbp]

\resizebox{\hsize}{!}{\includegraphics{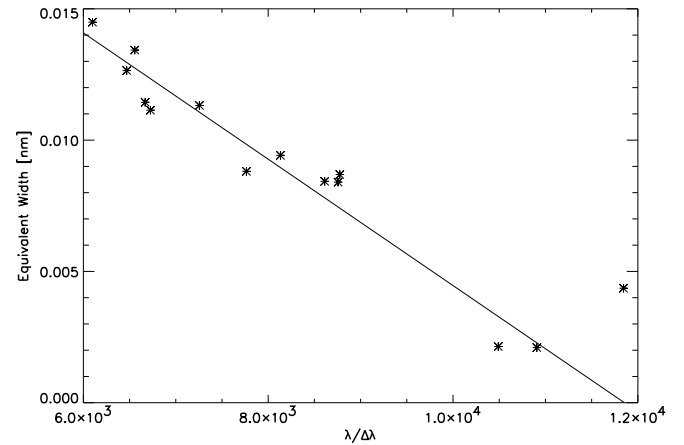}}

\caption{Equivalent widths of fitted lines plotted
against their observed $\lambda$/$\Delta$$\lambda$
values (Table\,1). The straight line is a least square
fit to the data points, except for absorption line no.
9, which was excluded in a conservative approach.
Extrapolating the linear fit to an equivalent width of
0 the observed line width should reproduce the
instrumental width. The values shown indicate an
instrumental resolution between 11\,000 and 12\,000.}

\label{fig7}
\end{figure}

Another approach to estimate the achieved instrumental
resolution is shown in Fig.~\ref{fig7}. The equivalent widths
of the fitted absorption lines are plotted against
their observed $\lambda$/$\Delta$$\lambda$ values. A
linear least square fit is applied to all data points
(point no.\,9 was excluded from the fit in a
conservative approach). Extrapolating this fitted
straight line to an equivalent width of 0 should
reproduce the instrumental resolution. The result is a
value between $\lambda$/$\Delta$$\lambda$\,=\,11\,000
and 12\,000.

This result is not really surprising, as the spectral
resolution of 10\,000 was calculated for a fully
illuminated 10{\arcsec} diaphragm. Though the Echelle
spectrometer was operated with a 20{\arcsec} diaphragm
the ASTRO-SPAS pointing jitter was $\pm$\,2{\arcsec}
and therefore the effective object image size was much
smaller than the assumed 10{\arcsec}.

\begin{figure*}[tbp]

\resizebox{12cm}{!}{\includegraphics{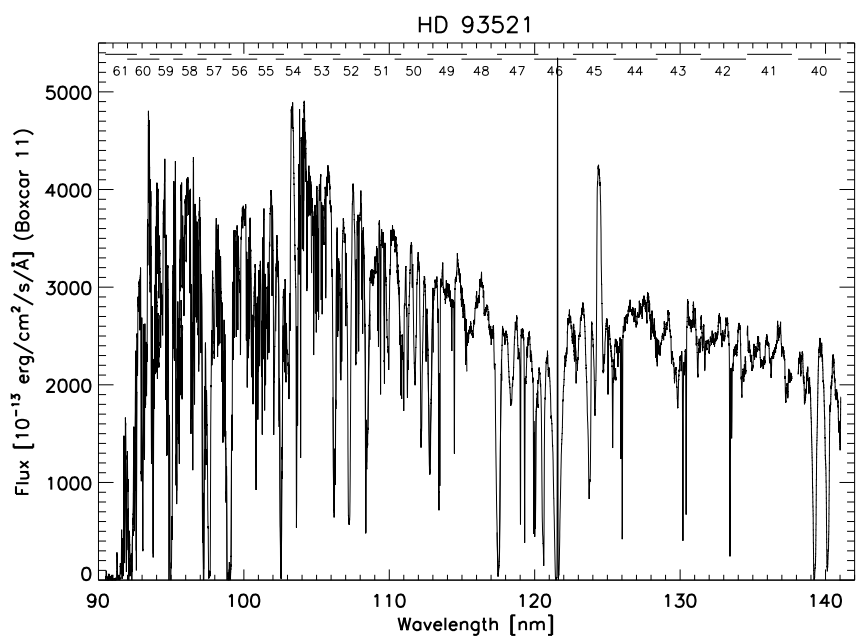}}
\hfill
\parbox[b]{55mm}{
\caption{ORFEUS\,II Echelle spectrum of HD\,93521. The
individual Echelle orders are plotted separately, the
wavelength range for each order is indicated at the
top of the figure. Except for orders 40/41 and 41/42
there is an overlap region between adjacent orders
which increases with the order number.
The spectrum is plotted with an 11 pixel wide boxcar
smoothing.}

\label{fig8}}
\end{figure*}

Figure \ref{fig8} shows the complete Echelle spectrum of
HD\,93521 smoothed with a 11 pixel wide boxcar. The
individual Echelle orders are plotted separately, the
spectral range for each order is indicated at the top
of the figure. The overlap region between adjacent
orders increases with the order number. A small gap
exists between orders 40 and 41 and between orders 41
and 42.

\section{Summary}

The Echelle spectrometer was one of the two focal
instruments of the ORFEUS telescope, which was flown
on its second mission in November/December 1996. The in-%
flight performance and the principles of data
reduction for this instrument are presented. The
wavelength range is 90\,nm to 140\,nm and we showed,
that the spectral resolution is significantly better
than $\lambda$/$\Delta$$\lambda$\,=\,10\,000, where
$\Delta$$\lambda$ is measured as the FWHM of the
instrumental profile. The effective area peaks at
1.3\,cm$^{2}$ near 110\,nm. The background is
dominated by straylight from the Echelle grating and
is about 15\% in an extracted spectrum for spectra
with a rather flat continuum. The internal accuracy of
the wavelength calibration is better than
$\pm$\,0.005\,nm. No corrections could be applied
which would correct for a possible systematic shift of
the object within the 20{\arcsec} aperture of the
telescope.

For the future we plan to improved the data quality by
further reduction steps and investigations:
\begin{itemize}
\item Integration of single photons with correction of
actual orbital radial velocity and possibly correction
for the ASTRO-SPAS pointing jitter.
\item Correlation of count rate variations with ASTRO-%
SPAS pointing jitter to estimate the object's position
within the aperture and the size and the shape of the
telescope's image of the object.
\item Investigation of small scale flat field
variations.
\item Investigation of the cause for the ``tilted''
absorption lines within the spectral orders.
\item Stray light model for improved background
subtraction.
\end{itemize}
\begin{acknowledgements}

This project could not have been successfully carried
out without the very efficient cooperation of all the
many individuals and institutions involved. In particular we
acknowledge the excellent performance of the teams of
DASA with the construction and mission operations of
the ASTRO-SPAS satellite, of Kayser-Threde with the
construction of the ORFEUS telescope, the ground
support teams at the Kennedy Space Center, the crew of
STS-80 {\it Columbia\/}, the mission representatives
of DARA and NASA and last not least our colleagues at
the Space Sciences Laboratory of the University of
California at Berkeley. We wish to thank Dr. Ralf
Napiwotzki for calculating the atmosphere models for
BD\,+28\degr\,4211 and making them available to us.
The ORFEUS program was supported by DARA grants
WE3\,OS\,8501 and WE2\,QV\,9304 and NASA grant NAG5-%
696.
\end{acknowledgements}

%\listofobjects
\end{document}